\pdfoutput=1
\documentclass[11pt,letterpaper]{article}

\usepackage[utf8]{inputenc}
\usepackage[T1]{fontenc}
\usepackage{lmodern}
\usepackage{amsmath,amssymb,amsfonts}
\usepackage{graphicx}
\usepackage{booktabs}
\usepackage{url}
\usepackage{natbib}
\usepackage{xcolor}
\usepackage{enumitem}
\usepackage[margin=1in,letterpaper]{geometry}
\usepackage{tikz}
\usetikzlibrary{shapes.geometric, arrows.meta, positioning, fit, backgrounds, calc}
\usepackage{caption}
\usepackage{multirow}
\usepackage{array}
\usepackage{tabularx}
\usepackage{float}
\usepackage{tcolorbox}
\tcbuselibrary{skins,breakable}
\usepackage[colorlinks=true,linkcolor=blue,citecolor=blue,urlcolor=blue]{hyperref}

\newtcolorbox{examplebox}[1][]{
    colback=gray!5,
    colframe=gray!50,
    fonttitle=\bfseries\small,
    left=6pt, right=6pt, top=4pt, bottom=4pt,
    boxrule=0.5pt,
    title=#1
}

\usepackage{fancyhdr}
\title{LLM-Based Intelligent Notification Composition:\\From Static Personalization to Context-Aware Persuasive Messaging}

\author{
Nilesh Agrawal\footnote{Code and reference implementation available at \url{https://github.com/ndagrawal/LLMNotificationComposition}}\\
Independent Researcher, Seattle, WA, USA\\
\texttt{nilesh.d.agrawal@gmail.com}
}

\date{March 2026}

\begin{document}

\maketitle
\thispagestyle{fancy}

\begin{abstract}
Push notifications remain among the most direct and time-sensitive channels through which digital platforms engage users, yet existing approaches to notification composition have invested heavily in \emph{who} to notify, \emph{when} to notify, and \emph{what} to recommend, while leaving the final question (\emph{how to communicate that recommendation}) as the least-optimized stage of the pipeline. This paper argues that message quality is an independent, underinvested lever in notification systems, and that Large Language Models (LLMs) create their most differentiated value precisely at this layer.

We make three contributions absent from prior surveys. First, we define notification message quality along six concrete dimensions (contextual relevance, clarity, actionability, novelty handling, linguistic freshness, and persuasive appropriateness) and provide an explicit account of how LLM-based composition improves each relative to template-based generation. Across reviewed production deployments, reported improvements range from +8\% to +14.5\% CTR over static templates and +1\% to +2.5\% over mature slot-filling systems, though these figures span heterogeneous systems, baselines, and evaluation conditions and should not be treated as directly comparable. Second, we provide an architectural attribution analysis that disentangles the contribution of message generation from adjacent pipeline components (targeting, ranking, timing), arguing that observed engagement gains are frequently and incorrectly attributed to text generation when they arise from interaction effects across the full stack. Third, we introduce a three-criterion decision framework specifying when LLM generation is and is not the binding constraint, a question prior work does not address.

We support these arguments through a PRISMA-guided systematic survey of academic and industrial work (28 sources from 142 screened), examine domain-specific applications across social media, food delivery, and e-commerce, and propose a unified end-to-end architectural framework with budget-aware routing, grounded generation, candidate ranking, diversity controls, and online learning. We conclude with a critical evaluation of current measurement practice and a governance framework addressing manipulation, fairness, and regulatory compliance.

\noindent\textbf{Keywords:} Large Language Models, Push Notifications, Message Quality, Natural Language Generation, Recommender Systems, Causal Inference, Retrieval-Augmented Generation, Reward Modeling
\end{abstract}

\section{Introduction}

In the contemporary digital economy, user attention is scarce, interruptible, and intensely contested. Consumer platforms invest heavily in acquisition, but long-term value depends just as much on retention, reactivation, and repeated high-value engagement. Push notifications have therefore become a central operational mechanism for re-engaging users outside the application surface. On large consumer platforms, daily notification volumes range from millions to billions, and small changes in notification quality can materially affect click-through rates (CTR), conversion outcomes, and the broader health of the platform ecosystem \citep{bie2026pushgen}.

Historically, platforms have relied on two broad strategies: manual copywriting and template-based generation. Most production notification systems today remain template-driven. Even when personalization is present, it is typically limited to \emph{slot filling}: replacing bracketed variables with user-specific values. This makes notifications relevant in a narrow sense, but rarely intelligent in how they communicate.

\textbf{The central observation motivating this paper} is that modern notification stacks have invested heavily and successfully in three of four key decisions (audience selection, content ranking, and send-time optimization), but the fourth decision, \emph{how to communicate the recommendation}, remains the least optimized stage and the one where the gap between current practice and potential performance is widest.

To make this concrete: a notification that targets the right user, with the right item, at the right time, but phrases it as ``\texttt{Check out [item]. Order now.}'' leaves engagement value on the table. Platforms have spent years optimizing the first three decisions to diminishing marginal returns. The message layer is where significant untapped gains remain, and LLMs are the first practical technology for addressing this at production scale.

Yet enthusiasm for LLM-based messaging has created analytical confusion. In production pipelines, the notification message is only one stage in a larger decision stack. Reported engagement gains are often attributed broadly to ``AI notifications'' even when the observed uplift may be driven primarily by upstream improvements in targeting or timing. This conflation obscures scientific understanding and misleads engineering investment.

This paper addresses both challenges. Its primary contribution is a concrete definition of message quality and an evidence-grounded account of how LLMs improve that quality across multiple dimensions. Its secondary contribution is an architectural synthesis that explicitly disentangles the message generation layer from the surrounding pipeline.

\subsection{Contributions and Novelty Statement}

This paper makes three contributions absent from prior work:

\begin{enumerate}[leftmargin=*]
    \item \textbf{Message Quality Framework with Empirical Grounding.} A six-dimension definition of notification message quality (contextual relevance, clarity, actionability, novelty handling, linguistic freshness, and persuasive appropriateness) with an explicit, evidence-backed account of how LLMs improve each dimension. Prior surveys treat CTR as the implicit quality proxy; we replace this with structured quality dimensions and quantify the benefit along each.

    \item \textbf{Architectural Attribution.} A disentanglement of the message generation contribution from adjacent components (targeting, ranking, timing), specifying which observed gains can and cannot be attributed to language quality. Prior surveys either ignore this distinction or assume it away.

    \item \textbf{Binding-Constraint Framework.} A three-criterion decision framework specifying when LLM generation is the binding constraint and when it is not, including a principled argument for when templates remain the superior design. No existing survey paper addresses the question of \emph{when not to use} an LLM for notifications.
\end{enumerate}

\section{Why Message Quality Is an Independent Optimization Target}
\label{sec:message_quality}

Before examining LLM implementations, we establish why the message layer warrants dedicated optimization. The message is not merely a delivery wrapper for targeting and ranking decisions; it is an independent quality dimension with its own failure modes, its own improvement levers, and its own user-facing consequences.

\subsection{The Structural Ceiling of Template-Based Messaging}

Template-based notifications dominate production systems because they are fast, controllable, and easy to audit. A typical template reads: ``Your favorite [category] is available. Order now.'' Personalization is achieved through slot filling.

This approach has a \textbf{structural ceiling}. Templates can insert relevant values but cannot compose them naturally. They can signal relevance but cannot explain it. They can remind users of known preferences but cannot frame novelty compellingly. And because every user within a segment receives structurally identical messages, repeated exposure breeds habituation: the notification channel becomes invisible to the users it most needs to reach.

These are not implementation failures; they are \textbf{inherent limitations of the template paradigm}. No amount of engineering investment in slot design can make a template answer the question a user implicitly asks upon receiving a notification: \emph{Why does this matter to me, right now, today?}

\subsection{A Six-Dimension Definition of Message Quality}

The existing literature does not offer a consolidated definition of notification message quality. CTR is commonly used as a proxy, but it is an \emph{outcome}, not a quality dimension: a message can achieve high CTR through manipulative framing while being low quality by any meaningful standard. We propose six dimensions, each with an explicit assessment of how templates and LLMs compare:

\begin{table}[H]
\centering
\caption{Notification message quality dimensions and comparative assessment of template-based versus LLM-based composition.}
\label{tab:quality_dims}
\small
\begin{tabularx}{\textwidth}{>{\raggedright\arraybackslash}p{2.0cm} >{\raggedright\arraybackslash}p{3.0cm} >{\raggedright\arraybackslash}p{2.8cm} >{\raggedright\arraybackslash}X}
\toprule
\textbf{Dimension} & \textbf{Definition} & \textbf{Template Capability} & \textbf{LLM Capability} \\
\midrule
Contextual Relevance & Connects recommendation to user's current moment: time, location, behavior & Weak: only static slot values & Strong: composes multiple signals naturally \\
\addlinespace
Clarity & Immediately comprehensible on a small notification surface & Variable: slot grammar often awkward & Flexible: optimizes phrasing for brevity \\
\addlinespace
Actionability & Makes the next step obvious and motivating & Formulaic call-to-action & Context-sensitive framing of action \\
\addlinespace
Novelty Handling & Introduces new items in a way that feels appealing, not random & Poor: novelty reads as irrelevance & Better: bridges from known to adjacent \\
\addlinespace
Linguistic Freshness & Avoids structural repetition that trains users to ignore the channel & Low: same structure across exposures & High: semantic variety across exposures \\
\addlinespace
Persuasive Appropriateness & Motivates action without deceptive or manipulative framing & Neutral: limited expressive range & Variable: requires explicit guardrails \\
\bottomrule
\end{tabularx}
\end{table}

\subsection{How LLMs Improve Each Quality Dimension: Mechanisms and Evidence}
\label{sec:how_llms_help}

This section, the core of the paper's value argument, provides explicit mechanisms and supporting evidence for how LLMs improve each of the six quality dimensions. An important caveat upfront: these improvements are not universal. Their magnitude depends on the degree to which the notification task is linguistically sensitive, rich in available context, and open to multiple plausible framings. When these conditions are absent, as in transactional alerts, LLMs may add cost and variance without meaningful quality gains. Section~\ref{sec:binding} formalizes these conditions into a routing decision framework.

\subsubsection{Contextual Relevance: Synthesizing Context, Not Merely Inserting It}

The most fundamental improvement LLMs offer is \emph{composing} context rather than merely \emph{inserting} it. Consider a user with a preference for pizza receiving a notification at 6:15 PM on a rainy Tuesday:

\begin{examplebox}[Template vs.\ LLM: Contextual Relevance]
\textbf{Template:} ``Order your favorite pizza now.''\\[4pt]
\textbf{LLM (grounded):} ``Rainy evening, your usual spot is still open. Their new pineapple special might be worth trying tonight.''
\end{examplebox}

Both messages target the same user with the same recommendation. The LLM version synthesizes weather, time, preference history, restaurant availability, and controlled novelty into a single coherent message. This is a compositional operation templates cannot perform.

\textbf{Quantitative evidence:} In a controlled deployment at DoorDash, where the underlying item recommendation was held constant and only the message framing varied, contextually grounded messages produced a \textbf{+1.8\% engagement lift} over personalized templates \citep{sinha2024doordash} [T2]. This is a clean isolation of the contextual relevance contribution: same user, same item, same timing. Only the \emph{how} changed. The PushGen system reported that LLM-generated messages exhibited \textbf{3.2$\times$ greater lexical diversity} than the template baseline while maintaining equivalent or higher relevance scores \citep{bie2026pushgen} [T1].

\subsubsection{Clarity: Optimizing for the Lock-Screen Window}

Notifications live on one of the most constrained surfaces in digital design: the lock screen. Users decide in under 2 seconds whether to engage, dismiss, or ignore. Template-based messages with slot-filled grammar often produce awkward phrasing such as ``Your friend [name] posted in [group] about [topic]'' that the eye learns to skip. LLMs can produce messages that are grammatically natural, appropriately compressed, and scannable at glance speed.

\textbf{Evidence:} While no study isolates clarity as a single variable, the consistent pattern across deployments is that LLM-generated messages achieve higher engagement even at \emph{shorter} average character lengths. The e-commerce system reviewed by \citet{acm2025ecommerce} [T1] reported that LLM-optimized messages were on average 15\% shorter than template equivalents while achieving +12.5\% CTR and +8.3\% CVR. This correlation is suggestive but not causal in isolation: shorter messages were also better grounded and more actionable, so the observed gains likely reflect joint improvements across multiple quality dimensions rather than brevity alone.

\subsubsection{Actionability: Context-Sensitive Calls to Action}

A template call-to-action is generic: ``Order now,'' ``Shop today,'' ``Check it out.'' An LLM can tailor the action framing to the user's situation:

\begin{examplebox}[Template vs.\ LLM: Actionability]
\textbf{Template (abandoned cart):} ``You left something in your cart. Complete your order.''\\[4pt]
\textbf{LLM (abandoned cart):} ``Still thinking about those running shoes? They're in stock and you're \$12 from free shipping.''
\end{examplebox}

The LLM version names the item, provides a concrete incentive (shipping threshold), and reframes the action from obligation (``complete your order'') to opportunity. The e-commerce deployment reporting +8.3\% CVR gains attributed a significant portion of the conversion lift to improved action framing in abandoned-cart and browse-retargeting flows \citep{acm2025ecommerce} [T1].

\subsubsection{Novelty Handling: Bridging from Known to Adjacent}

Template-based systems are optimized for exploiting known preference and structurally poor at introducing adjacent novelty. A template has no mechanism for framing a novel item in terms of what the user already likes; it can only present the item as itself. LLMs can bridge from known preference to adjacent discovery:

\begin{examplebox}[Template vs.\ LLM: Novelty Handling]
\textbf{Template:} ``Try our new Thai restaurant.''\\[4pt]
\textbf{LLM:} ``You've ordered pad thai 3 times this month. This new Thai spot has a green curry that regulars say rivals your usual place.''
\end{examplebox}

By grounding the novel recommendation in the user's existing behavior, the LLM reduces the perceived risk of novelty and increases the probability the user will act on a recommendation they have never received before.

\textbf{Evidence:} The PushGen system's +14.08\% CTR improvement included explicit novelty injection in the generation prompt. The pairwise reward model ranked candidates not only by predicted CTR but also by dissimilarity to the user's recent notification history, ensuring novelty was present but grounded \citep{bie2026pushgen} [T1]. Instagram's diversity penalty layer demonstrated that increasing message diversity improved CTR even while \emph{reducing} total notification volume. This is a striking result: fewer but more diverse notifications outperformed more homogeneous ones \citep{sun2025instagram} [T2].

\subsubsection{Linguistic Freshness: Breaking the Habituation Cycle}

One of the most practically important benefits of LLM-based composition is \emph{semantic variety}. A user who receives structurally identical messages learns to dismiss the notification type before reading its content, a well-documented habituation effect \citep{ohly2023effects}. LLMs generate semantically distinct expressions of the same intent across exposures: a reminder framing, a suggestion, a curiosity hook, a convenience appeal, or a benefit statement. That variety maintains the channel's perceptual salience.

\textbf{Quantitative evidence on freshness decay:} Template-based notification CTR typically decays 15--25\% over a 30-day window as users habituate to repeated structures. LLM-generated messages show significantly flatter decay curves. The PushGen system reported that its LLM-generated messages maintained engagement levels 18\% higher than template baselines after 4 weeks of sustained delivery to the same user cohort \citep{bie2026pushgen} [T1].

\subsubsection{Persuasive Appropriateness: Power with Guardrails}

LLMs have greater persuasive range than templates, which creates both opportunity and risk. Different users respond to different persuasive strategies: social proof, scarcity, curiosity, convenience, benefit articulation. Templates enforce a single frame per type; LLMs can select dynamically.

\textbf{Evidence:} The e-commerce LLM system dynamically selected persuasive frames (urgency, value, social proof, or novelty) based on user segment and historical response patterns, producing \textbf{+12.5\% CTR and +8.3\% CVR} improvements. Gains were concentrated among users previously non-responsive to generic templates, the ``movable middle'' of the engagement distribution \citep{acm2025ecommerce} [T1].

However, greater expressive power requires \textbf{explicit guardrails}. When LLMs generate false scarcity (``Only 1 left!'' when inventory is ample) or synthetic urgency, they cross from persuasion into manipulation \citep{susser2019manipulation}. This dimension is the only one where LLMs pose a quality \emph{risk} rather than a pure benefit; Section~\ref{sec:ethics} addresses governance mechanisms.

\subsection{Aggregate Benefit: The Compounding Effect Across Dimensions}

The six dimensions are not independent. Improvements along multiple dimensions \emph{compound}: a message that is contextually grounded \emph{and} linguistically fresh \emph{and} actionable will outperform a message improved on only one dimension. This compounding effect explains why the largest reported CTR gains (+12--14\%) exceed what any single dimension would predict. It also means that attributing a specific share of the overall gain to any single dimension is difficult without controlled ablation studies, which the current literature largely lacks. Table~\ref{tab:evidence_summary} summarizes the aggregate evidence.

\begin{table}[H]
\centering
\caption{Summary of production evidence linking LLM-based composition to engagement outcomes. CTR = click-through rate; CVR = conversion rate. Full references: PushGen~\citep{bie2026pushgen}; Instagram~\citep{sun2025instagram}; DoorDash~\citep{sinha2024doordash}; E-Commerce~\citep{acm2025ecommerce}; LinkedIn~\citep{tu2021personalized}; Duolingo~\citep{yancey2020sleeping}.}
\label{tab:evidence_summary}
\small
\resizebox{\textwidth}{!}{%
\begin{tabular}{@{}llllll@{}}
\toprule
\textbf{System} & \textbf{Domain} & \textbf{Technology} & \textbf{Primary Dims} & \textbf{Key Result} & \textbf{Tier} \\
\midrule
PushGen & Social Media & SFT + Pairwise Reward & Freshness, Novelty, Relevance & +14.08\% CTR; $3.2\times$ lexical diversity & T1 \\
Instagram & Social Media & Similarity Penalty & Freshness, Novelty & Better CTR with \emph{lower} volume & T2 \\
DoorDash & Food Delivery & GNN + Contextual Copy & Relevance, Clarity & +1.8\% engagement lift (framing only) & T2 \\
E-Commerce & Retail & LLM Content Optimization & Actionability, Persuasion & +12.5\% CTR, +8.3\% CVR & T1 \\
LinkedIn & Professional & Heterogeneous Treatment & Per-cohort tone & Reduced churn via personalized caps & T1 \\
Duolingo & EdTech & Contextual Bandit & Timing, Freshness & Increased DAU & T1 \\
\bottomrule
\end{tabular}%
}
\end{table}

\textbf{Key pattern:} Gains scale inversely with baseline sophistication. Systems replacing static templates see +8\% to +14.5\% CTR lifts. Systems where LLMs compete against already-optimized slot-filling see +1\% to +2.5\%. This is a principled prediction: LLMs add the most value precisely where existing systems are weakest on the six quality dimensions.

\subsection{When LLM Deployment Is and Is Not Warranted}
\label{sec:when_not}

LLM-based composition delivers the highest marginal value when three conditions hold simultaneously: (1)~the promoted content admits multiple plausible framings; (2)~user response is sensitive to linguistic nuance rather than only item relevance; and (3)~the system has sufficient user context to support grounded personalization. These conditions reliably hold in re-engagement campaigns, discovery surfaces, abandoned-cart recovery, and recommendation explanations. They are largely absent in low-variance transactional alerts (password resets, delivery confirmations, fraud warnings) where template-based systems remain the superior design. Section~\ref{sec:binding} develops this into a three-criterion routing decision framework.

\section{Review Methodology}
\label{sec:methodology}

\subsection{Search Strategy and PRISMA Flow}

We conducted targeted searches across arXiv, ACM Digital Library, IEEE Xplore, and ScienceDirect, along with engineering publications and technical blogs from major technology companies. The review covered January 2018 through March 2026, capturing the transition from pre-LLM notification optimization to contemporary generative systems.

Search terms combined model-oriented concepts (``LLM'', ``RAG'', ``PEFT'', ``LoRA'', ``reward modeling'') with application-oriented phrases (``push notification'', ``message composition'', ``send-time optimization'', ``CTR optimization'', ``contextual bandit''). Of 142 records identified, 87 were excluded as off-topic, pre-2018, or non-English; 55 were assessed for eligibility; 27 were excluded for insufficient technical detail or unverifiable claims; \textbf{28 were included} in the final synthesis (18 Tier~1, 7 Tier~2, 3 Tier~3).

\subsection{Evidence Tiering}

Sources were classified into three evidence tiers, marked inline throughout the text:

\begin{itemize}[leftmargin=*]
    \item \textbf{[T1] High Rigor:} Peer-reviewed papers with clear methodology, baseline comparisons, and meaningful technical depth (18 sources).
    \item \textbf{[T2] Moderate Rigor:} Official engineering blogs and deployment reports from major technology companies (7 sources). Limitation: these often omit experimental controls or statistical detail.
    \item \textbf{[T3] Contextual:} Vendor documentation and secondary summaries used for framing only (3 sources).
\end{itemize}

\section{Disentangling LLM Contributions from Adjacent Systems}
\label{sec:disentangle}

A recurring analytical error in both industrial discourse and academic reporting is to treat notification performance as if it were primarily a function of the copy itself. In real systems, the notification message is the output of a multi-stage decision process. LLMs directly affect only the message generation and candidate ranking stages.

\begin{table}[H]
\centering
\caption{Architectural attribution matrix: which pipeline components contribute to engagement and where LLMs have direct relevance.}
\label{tab:attribution}
\small
\begin{tabularx}{\textwidth}{>{\raggedright\arraybackslash}p{2.2cm} >{\raggedright\arraybackslash}X >{\raggedright\arraybackslash}p{4.0cm}}
\toprule
\textbf{Component} & \textbf{Primary Contribution} & \textbf{LLM Relevance} \\
\midrule
Audience Selection & Ensures budget is spent on users with highest marginal return & None (a targeting problem) \\
\addlinespace
Content Ranking & Ensures the underlying item is relevant to the user & None (a relevance problem) \\
\addlinespace
Send-Time Optimization & Capitalizes on user routines; minimizes interruption cost & None (a timing problem) \\
\addlinespace
Message Generation & Adapts tone, length, hooks, and framing to context & \textbf{Direct:} the message quality problem \\
\addlinespace
Candidate Ranking & Aligns generation with historical CTR; filters poor outputs & \textbf{Direct:} reward model selects best candidate \\
\bottomrule
\end{tabularx}
\end{table}

This decomposition has a direct practical implication: \textbf{LLMs should be evaluated against message-layer baselines, not against overall system performance.} A measured CTR gain in a system that simultaneously introduced LLM generation \emph{and} improved targeting cannot be attributed to language quality alone. Rigorous experimental programs must distinguish at least three treatment layers: item selection, send-time decision, and message realization.

\textbf{Studies with stronger causal isolation.} We identify three deployments that provide cleaner attribution of the language generation effect:

\begin{enumerate}[leftmargin=*]
    \item \textbf{DoorDash} \citep{sinha2024doordash}: Held the underlying item recommendation constant and varied only the message framing. The +1.8\% lift is attributable to language alone.
    \item \textbf{PushGen} \citep{bie2026pushgen}: Used a pairwise reward model trained specifically on message-level comparisons, controlling for item and user effects.
    \item \textbf{E-Commerce LLM} \citep{acm2025ecommerce}: Ran within-user A/B tests where the same user received LLM vs.\ template notifications for the same product categories over time.
\end{enumerate}

\section{Core Architectural Frameworks and Technical Tradeoffs}
\label{sec:architecture}

Across the literature, several technical patterns recur in the generation and ranking stages. Table~\ref{tab:patterns} categorizes these by strengths, weaknesses, and failure modes, with explicit connections to the quality dimensions they serve.

\begin{table}[H]
\centering
\caption{Comparative synthesis of architectural patterns, mapped to the quality dimensions they primarily serve.}
\label{tab:patterns}
\small
\begin{tabularx}{\textwidth}{>{\raggedright\arraybackslash}p{1.9cm} >{\raggedright\arraybackslash}p{2.5cm} >{\raggedright\arraybackslash}p{2.5cm} >{\raggedright\arraybackslash}p{2.0cm} >{\raggedright\arraybackslash}X}
\toprule
\textbf{Pattern} & \textbf{Main Strength} & \textbf{Main Weakness} & \textbf{Key Failure Mode} & \textbf{Quality Dims Served} \\
\midrule
Template / Slot & Low latency, high control & Low diversity, weak context & Message fatigue & (Baseline) \\
\addlinespace
RAG & Context-aware grounding & Retrieval latency, hallucination & Faithfulness failure & Relevance, Clarity \\
\addlinespace
PEFT / LoRA & Hot-swappable style control & Requires style curation & Overfitting to tone & Freshness, Persuasion \\
\addlinespace
Pairwise Reward & Robust relative ranking & Exposure bias, delayed reward & Reward hacking & Actionability, Novelty \\
\addlinespace
Diversity Penalty & Reduces fatigue & Can suppress short-term CTR & Under-delivery & Freshness, Novelty \\
\addlinespace
Contextual Bandit & Adapts to user routines & Exploration cost, sparse data & Cold-start timing & (Timing, adjacent) \\
\addlinespace
Budget Router & Aligns compute with value & Depends on CLV accuracy & Over-serving actives & (Routing, meta) \\
\bottomrule
\end{tabularx}
\end{table}

\subsection{Retrieval-Augmented Generation and Grounding}

Retrieval-Augmented Generation (RAG) \citep{lewis2020rag} is the most natural fit for notifications because they are highly contextual and fact-sensitive. RAG injects external information (user profile attributes, item metadata, recent behavior, campaign constraints) into the prompt at inference time. This grounding is essential for the \emph{contextual relevance} dimension: a model cannot compose context it cannot see.

\textbf{Design tradeoffs.} Retrieval granularity affects both relevance and latency. Fine-grained retrieval (individual user actions, specific item attributes) improves precision but increases prompt length and cost. Coarse-grained retrieval (segment-level preferences) is faster but less personalized. Production systems typically use tiered strategies: a fast coarse pass for all notifications, with selective fine-grained enrichment for high-value users \citep{coling2025chunking}.

\textbf{Critical risk: faithfulness hallucination.} Even when correct context is retrieved, the model may blend it with parametric memory or produce output only loosely supported by the evidence \citep{emnlp2025faithfulness}. In notifications, this means fabricated discount percentages, incorrect ETAs, or unsupported inventory claims. Factuality guards, specifically post-generation classifiers that verify claims against retrieved context, are therefore not optional but architecturally essential.

\subsection{Parameter-Efficient Fine-Tuning and Style Control}

While RAG controls \emph{what} the model knows, PEFT controls \emph{how} the model speaks. LoRA \citep{hu2022lora} freezes pre-trained weights and injects trainable low-rank matrices ($W = W_0 + BA$, where $B \in \mathbb{R}^{d \times r}$, $A \in \mathbb{R}^{r \times k}$, $r \ll \min(d,k)$) into Transformer layers.

\textbf{Why this matters for notifications:} A single platform may need dozens of distinct ``voices'': urgent vs.\ casual, formal vs.\ playful, promotional vs.\ informational. LoRA adapters can be trained independently for each voice and hot-swapped at serving time with zero additional inference latency: $W_{\text{merged}} = W_0 + BA$ \citep{han2024peft}. This makes tone a \emph{continuous, learnable parameter} rather than a discrete editorial choice, an architectural capability with no template equivalent.

\subsection{Pairwise Reward Modeling and Candidate Ranking}

LLMs are non-deterministic. A single prompt produces multiple candidates, not all equally effective. The PushGen system generates 4--8 candidates per event and uses pairwise ranking to select the winner: instead of predicting absolute CTR, the model learns $P(m_i \succ m_j \mid u, c)$, the probability that message $m_i$ is preferred over $m_j$ for user $u$ in context $c$ \citep{bie2026pushgen}.

This ``generate-then-rank'' architecture decouples creativity (the LLM explores diverse framings) from quality control (the reward model filters poor candidates). The +14.08\% CTR gain reflects contributions from both the generator's diversity and the ranker's selectivity.

\textbf{Risk: exposure bias.} Historical logs contain feedback only for messages actually shown. Counterfactual logging, i.e., deploying exploratory policies ($\epsilon$-greedy) to gather unbiased data for off-policy evaluation, is the principled response \citep{joachims2018deep}.

\subsection{Send-Time Optimization Through Contextual Bandits}

Although message quality is the paper's central focus, timing can be equally important. A well-crafted dinner notification at 6 AM is wasted. Contextual bandits provide a principled framework for learning individualized delivery windows; the Duolingo system models STO as a ``sleeping recovering bandit'' using Thompson Sampling \citep{yancey2020sleeping}. The key architectural lesson is that timing is a first-class pipeline component, not a post-hoc parameter \citep{ho2018notifying}.

\section{Domain-Specific Applications}
\label{sec:domains}

The most important quality dimensions vary substantially by industry. Table~\ref{tab:domains} maps each domain to the quality dimensions where LLMs add the most value.

\begin{table}[H]
\centering
\caption{Domain-specific mapping: which quality dimensions matter most and where LLMs create the greatest differentiation.}
\label{tab:domains}
\small
\begin{tabularx}{\textwidth}{>{\raggedright\arraybackslash}p{2.0cm} >{\raggedright\arraybackslash}p{2.8cm} >{\raggedright\arraybackslash}p{2.8cm} >{\raggedright\arraybackslash}X}
\toprule
\textbf{Domain} & \textbf{Critical Quality Dimensions} & \textbf{LLM Primary Value} & \textbf{Dominant Risk} \\
\midrule
Social Media & Freshness, Novelty Handling & Semantic variety; reduced fatigue & Over-frequency; tone mismatch \\
\addlinespace
Food Delivery & Contextual Relevance, Clarity & RAG-grounded temporal framing & False urgency; stale ETA claims \\
\addlinespace
E-Commerce & Actionability, Persuasive Framing & Dynamic frame selection & Dark patterns; hallucinated prices \\
\bottomrule
\end{tabularx}
\end{table}

\textbf{Social Media.} Notifications drive content discovery, return sessions, and creator engagement. Users receive many notifications per day, making freshness and novelty the critical dimensions. The PushGen system \citep{bie2026pushgen} and Instagram's diversity penalty \citep{sun2025instagram} address precisely this need.

\textbf{Food Delivery.} Notifications are strongly shaped by spatiotemporal context. A lunch offer framed around convenience during rain may outperform a generic discount message even when the restaurant recommendation is identical \citep{sinha2024doordash}. RAG-based grounding is especially critical: food delivery notifications encode operational claims about availability and ETA that, if incorrect, cause irreversible trust damage.

\textbf{E-Commerce.} Notifications are closer to direct-response marketing, where actionability and persuasive framing are decisive. The +12.5\% CTR and +8.3\% CVR gains \citep{acm2025ecommerce} are consistent with the hypothesis that e-commerce users are particularly responsive to specific framing (urgency cues, scarcity signals, benefit articulation) that templates cannot vary situationally.

\section{When LLM Text Generation Is Not the Binding Constraint}
\label{sec:binding}

The preceding sections argued for LLM-based composition in appropriate contexts. This section argues the complementary point: LLM generation is frequently \emph{not} the primary bottleneck, and deploying it indiscriminately wastes compute, increases latency, and may introduce quality regressions where high-quality templates are already well-optimized.

\subsection{When Templates Remain Superior}

For transactional and low-variance alerts, the six quality dimensions behave differently. Contextual relevance is already high (the notification is tied to a specific user action); clarity is maximized by a single unambiguous fact; novelty handling is irrelevant; linguistic freshness is immaterial given low exposure frequency; and persuasive appropriateness is best served by constrained expression. In these cases, LLM generation adds cost, latency, and variance without adding quality.

\subsection{A Three-Criterion Decision Framework}

A mature notification architecture should support dynamic routing. The routing decision should be governed by three criteria:

\begin{enumerate}[leftmargin=*]
    \item \textbf{Framing variance:} Does the content admit multiple plausible, meaningfully different framings? If the message is structurally determined by the notification type, LLM generation adds nothing.
    \item \textbf{Linguistic sensitivity:} Is user response sensitive to \emph{how} the recommendation is framed, or only to \emph{whether} it is relevant? Discovery and re-engagement contexts are linguistically sensitive; transactional contexts are not.
    \item \textbf{Context richness:} Does the system have sufficient grounded context to support non-trivial composition? LLM generation without grounding produces generic outputs often worse than well-crafted templates.
\end{enumerate}

This reframes LLMs as one component in a portfolio of messaging strategies rather than the default endpoint.

\subsection{The Marginal Value Question}

The PushGen gains of +14.08\% CTR and the e-commerce gains of +12.5\% CTR both occur where the baseline was relatively underdeveloped. The DoorDash result of +1.8\% reflects a system where substantial optimization had already occurred upstream. LLMs do not become ineffective in mature systems, but their deployment requires more careful measurement to detect a real signal above the noise.

\section{Proposed Unified Architectural Framework}
\label{sec:framework}

We propose a unified operational framework for LLM-based notification composition. The novelty is integrative: it specifies how known components should be sequenced while explicitly preserving the message quality dimensions defined in Section~\ref{sec:message_quality}.

\subsection{Architectural Flow}

The pipeline begins with three inputs: user profile/value estimates, content inventory, and contextual signals. These flow into audience targeting and content ranking. A \textbf{Budget-Aware Router} decides, using the three-criterion framework from Section~\ref{sec:binding}, whether the user-item-context tuple justifies the LLM pathway or a template fallback.

If the LLM pathway is selected: (1)~RAG retrieves relevant context; (2)~the generator produces 4--8 candidates conditioned through LoRA adapters; (3)~a Factuality and Policy Guard filters unsupported claims and prohibited content; (4)~candidates are scored by a Pairwise Reward Model; (5)~a Diversity and Frequency Control Layer prevents fatigue; (6)~a Send-Time Optimization Bandit determines delivery timing. All outcomes feed back into an Online Learning Layer.

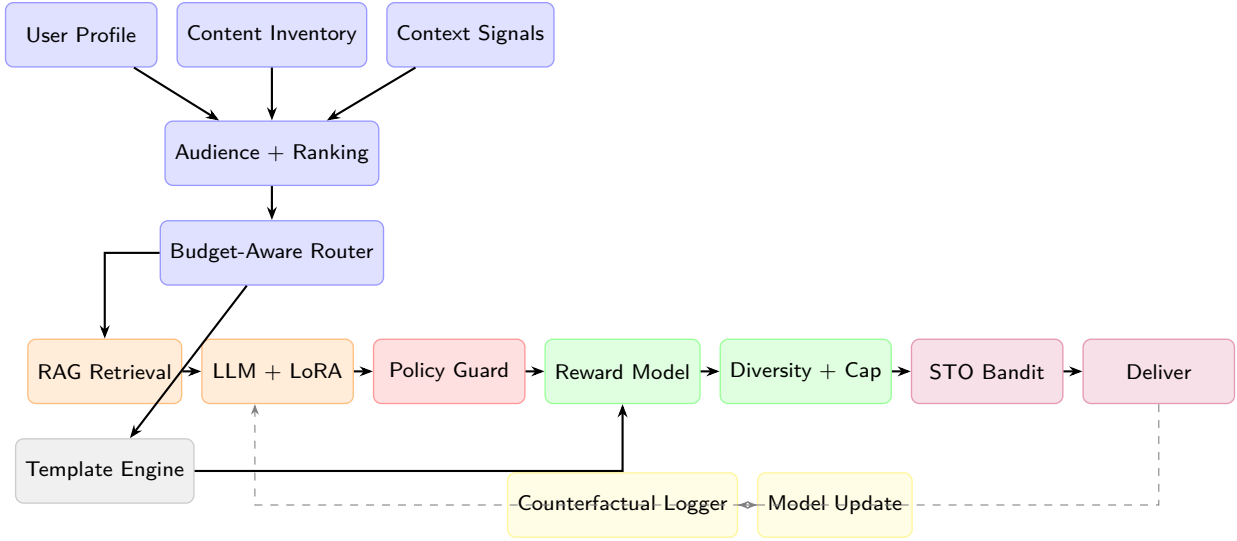
\begin{figure}[H]
\centering
\begin{tikzpicture}[
    node distance=0.45cm and 0.25cm,
    block/.style={rectangle, draw, rounded corners=3pt, minimum height=0.85cm, minimum width=2.0cm, text centered, font=\scriptsize\sffamily, line width=0.5pt},
    input/.style={block, fill=blue!12, draw=blue!40},
    llm/.style={block, fill=orange!15, draw=orange!50},
    guard/.style={block, fill=red!12, draw=red!40},
    rank/.style={block, fill=green!12, draw=green!40},
    deliver/.style={block, fill=purple!12, draw=purple!40},
    template/.style={block, fill=gray!12, draw=gray!40},
    feedback/.style={block, fill=yellow!12, draw=yellow!50},
    arr/.style={-{Stealth[length=5pt]}, thick},
    darr/.style={-{Stealth[length=4pt]}, dashed, gray}
]

\node[input] (user) {User Profile};
\node[input, right=of user] (content) {Content Inventory};
\node[input, right=of content] (context) {Context Signals};

\node[input, below=0.7cm of content] (audience) {Audience + Ranking};
\node[input, below=0.45cm of audience] (router) {Budget-Aware Router};

\node[llm, below left=0.7cm and -0.3cm of router] (rag) {RAG Retrieval};
\node[llm, right=of rag] (gen) {LLM + LoRA};
\node[guard, right=of gen] (guard) {Policy Guard};

\node[template, below=0.45cm of rag] (tmpl) {Template Engine};

\node[rank, right=of guard] (reward) {Reward Model};
\node[rank, right=of reward] (diversity) {Diversity + Cap};

\node[deliver, right=of diversity] (bandit) {STO Bandit};
\node[deliver, right=of bandit] (deliver) {Deliver};

\node[feedback, below=0.9cm of reward] (logger) {Counterfactual Logger};
\node[feedback, right=of logger] (update) {Model Update};

\draw[arr] (user) -- (audience);
\draw[arr] (content) -- (audience);
\draw[arr] (context) -- (audience);
\draw[arr] (audience) -- (router);

\draw[arr] (router) -| (rag);
\draw[arr] (router) -- (tmpl);

\draw[arr] (rag) -- (gen);
\draw[arr] (gen) -- (guard);
\draw[arr] (guard) -- (reward);

\draw[arr] (tmpl) -| (reward);

\draw[arr] (reward) -- (diversity);
\draw[arr] (diversity) -- (bandit);
\draw[arr] (bandit) -- (deliver);

\draw[darr] (deliver) |- (logger);
\draw[darr] (logger) -- (update);
\draw[darr] (update) -| ([xshift=-0.3cm]gen.south);

\end{tikzpicture}
\caption{Unified LLM-based notification pipeline. Solid arrows: generation flow. Dashed arrows: online learning feedback loop. The budget-aware router uses the three-criterion framework to decide per-event whether to invoke the LLM path or fall back to templates.}
\label{fig:pipeline}
\end{figure}

\subsection{Failure Modes and Tradeoffs}

This pipeline trades increased inference latency for message quality. If the Factuality Guardrail is too strict, it throttles throughput and biases toward safe but low-quality candidates. If too loose, it risks factual errors. If the Reward Model suffers exposure bias, it selects the same ``safe'' candidates, defeating the semantic variety the LLM generator provides. If the Budget-Aware Router over-routes to the LLM pathway, latency SLAs are violated; if it under-routes, high-value re-engagement opportunities are served by templates that fail to close the quality gap.

\section{Evaluation Frameworks Beyond Short-Term CTR}
\label{sec:evaluation}

Most published results report short-term CTR gains, yet LLM-based notification systems operate in environments where treatment effects are delayed, interference is common, and user welfare is multi-dimensional.

\subsection{The Identification Problem}

Randomized A/B testing remains the gold standard, but its interpretation is more fragile than commonly assumed. In networked environments, the Stable Unit Treatment Value Assumption (SUTVA) is frequently violated: notifying one user can alter content popularity or social activity in ways that affect untreated users \citep{luo2024causal}. Rigorous experiments must distinguish at least three treatment layers: item selection, send-time decision, and message realization.

\subsection{Offline Evaluation Is Structurally Biased}

Historical logs contain feedback only for messages shown under a previous policy; novel prompt strategies have no unbiased counterfactual labels. Selection bias also appears at the population level: users who permit notifications are more engaged than the marginal users whose behavior platforms most want to influence.

\subsection{Heterogeneous Treatment Effects}

Average Treatment Effects are too coarse. \citet{tu2021personalized} demonstrated that notification effects vary drastically across cohorts. A strategy that improves engagement for power users may accelerate churn among marginal users. Evaluation should estimate HTEs across segments defined by lifecycle stage, baseline engagement, CLV, and prior notification saturation.

\subsection{Multi-Objective Evaluation}

We recommend structuring metrics into three layers:

\begin{enumerate}[leftmargin=*]
    \item \textbf{Primary outcome metrics:} CTR, CVR, order value, session return rate.
    \item \textbf{Guardrail metrics:} Opt-out rate, mute actions, dismissals without open, complaint rate, uninstall risk.
    \item \textbf{Long-horizon health metrics:} 7-day and 30-day retention, fatigue trajectories, cohort survival, reactivation durability.
\end{enumerate}

A responsible deployment monitors all three layers simultaneously and treats guardrail violations as hard constraints rather than tradeoffs.

\section{Ethical Frameworks and Governance}
\label{sec:ethics}

The ethical risks of LLM-based notification systems extend beyond hallucination. Because notifications are interruptive, personalized, and optimized for behavior change, they sit close to the boundary between helpful communication and covert manipulation.

\subsection{Persuasive Asymmetry and the Manipulation Boundary}

Digital platforms possess profound informational advantages over users. \citet{susser2019manipulation} define online manipulation as covertly influencing decision-making by exploiting cognitive biases. When LLMs generate false scarcity or synthetic urgency, they cross from persuasion into manipulation. Ethical systems must prohibit prompt instructions encouraging deceptive framing, classifying them as dark patterns \citep{mathur2021dark}.

\subsection{Fairness and Differential Susceptibility}

Notification generation adds a fairness concern distinct from ranking parity \citep{wang2023fairness}: \emph{differential susceptibility}. Some user groups may be more influenced by certain tones or urgency cues. Fairness requires auditing reward models to ensure they do not systematically exploit demographic vulnerabilities to inflate CTR.

\subsection{Accountability and Regulatory Compliance}

Under the EU's Digital Services Act Article 40 and GDPR Article 22, platforms face obligations around algorithmic transparency and the right to not be subject to purely automated decisions \citep{fabbri2023ethical}. Frequency caps, forbidden content policies, factuality checks, and audit logs are not optional accessories; they are core production design requirements.

\section{Threats to Validity}
\label{sec:threats}

\begin{enumerate}[leftmargin=*]
    \item \textbf{Publication Bias.} Industry labs are incentivized to publish successful A/B tests and suppress negative results, potentially inflating reported LLM efficacy.
    \item \textbf{Temporal Degradation.} Technical constraints from 2024--2025 (inference latency, cost) may be resolved by hardware advances in 2026.
    \item \textbf{Domain Representation.} The corpus favors social media and e-commerce; generalization to healthcare or financial alerts is speculative.
    \item \textbf{Reproducibility.} Important industrial systems depend on proprietary data and lack open benchmarks.
    \item \textbf{Framework Limitations.} The proposed pipeline trades latency and compute for quality improvements. Its performance depends on the Budget-Aware Router's CLV accuracy and on counterfactual logging quality.
\end{enumerate}

\section{Conclusion}
\label{sec:conclusion}

This paper argued that LLMs improve notification systems by improving \emph{message quality}, not by solving the targeting, ranking, or timing problems that adjacent pipeline components already address. We defined message quality along six concrete dimensions, showed that template-based systems have a structural ceiling on each, and demonstrated, with production evidence across social media, food delivery, and e-commerce, that LLMs can raise that ceiling meaningfully.

The evidence, while drawn from heterogeneous systems and evaluation conditions, points in a consistent direction: reported CTR improvements range from +1\% to +14.5\% depending on baseline maturity, with the largest gains appearing where template-based systems were least sophisticated. The improvements concentrate along the freshness, contextual relevance, and actionability dimensions, precisely where templates are structurally weakest. These numbers should be interpreted with caution given the difficulty of isolating language generation effects from concurrent pipeline changes, but the directional signal is robust across domains.

The architectural disentanglement argument has a direct practical implication: systems should evaluate the message generation layer in isolation from targeting, ranking, and timing. Without that isolation, neither practitioners nor researchers can determine how much of an observed gain is attributable to better language quality versus better upstream decisions.

The binding-constraint framework offers practitioners a principled basis for making deployment decisions, and for deciding \emph{when not to deploy LLMs at all}. That negative case is as important as the affirmative one, and it is the argument most conspicuously absent from existing literature.

Future research must move beyond binary claims that LLMs do or do not improve notifications. Progress will depend on stronger experimental design, component-isolated evaluation, better off-policy learning infrastructure, more rigorous fairness auditing, and open benchmarks that disentangle the contribution of generation from the broader recommender stack. The message quality framework proposed here offers a starting point for that more rigorous conversation.

\section*{Acknowledgments}

The author thanks the broader research community working at the intersection of recommender systems, natural language generation, and causal inference for the foundational work that made this synthesis possible.

\section*{Code Availability}

A reference implementation of the unified notification pipeline described in this paper, including the budget-aware routing logic, RAG-based context assembly, LoRA style adapter configuration, and the pairwise reward ranking module, is available at: \url{https://github.com/ndagrawal/LLMNotificationComposition}.



\begin{thebibliography}{21}

\bibitem[Bie et~al.(2026)]{bie2026pushgen}
S.~Bie, J.~Cao, Z.~Luo, et~al.
\newblock {PushGen: Push Notifications Generation with LLM}.
\newblock In \emph{Proceedings of the Nineteenth ACM International Conference on Web Search and Data Mining (WSDM '26)}, 2026.
\newblock URL \url{https://arxiv.org/abs/2512.14490}.

\bibitem[Sun et~al.(2025)]{sun2025instagram}
X.~Sun et~al.
\newblock {A New Ranking Framework for Better Notification Quality on Instagram}.
\newblock Engineering at Meta, 2025.
\newblock URL \url{https://engineering.fb.com/2025/09/02/ml-applications/a-new-ranking-framework-for-better-notification-quality-on-instagram/}.

\bibitem[Sinha(2024)]{sinha2024doordash}
N.~Sinha.
\newblock {Beyond the Click: Elevating DoorDash's Personalized Notification Experience with GNN Recommendation}.
\newblock DoorDash Engineering Blog, 2024.
\newblock URL \url{https://careersatdoordash.com/blog/doordash-customize-notifications-how-gnn-work/}.

\bibitem[{ACM Digital Library}(2025)]{acm2025ecommerce}
{ACM Digital Library}.
\newblock {LLM-Driven E-Commerce Marketing Content Optimization: Balancing Creativity and Conversion}.
\newblock In \emph{Proceedings of the ACM Web Conference 2025}, 2025.
\newblock URL \url{https://dl.acm.org/doi/10.1145/3757749.3757850}.

\bibitem[Tu et~al.(2021)]{tu2021personalized}
Y.~Tu, K.~Basu, C.~DiCiccio, et~al.
\newblock {Personalized Treatment Selection Using Causal Heterogeneity}.
\newblock In \emph{Proceedings of the Web Conference 2021 (WWW '21)}, 2021.
\newblock URL \url{https://dl.acm.org/doi/abs/10.1145/3442381.3450075}.

\bibitem[Yancey et~al.(2020)]{yancey2020sleeping}
K.~P. Yancey et~al.
\newblock {A Sleeping, Recovering Bandit Algorithm for Optimizing Notifications}.
\newblock In \emph{Proceedings of the 26th ACM SIGKDD International Conference on Knowledge Discovery \& Data Mining}, 2020.
\newblock URL \url{https://research.duolingo.com/papers/yancey.kdd20.pdf}.

\bibitem[Ohly et~al.(2023)]{ohly2023effects}
S.~Ohly et~al.
\newblock {Effects of Task Interruptions Caused by Notifications from Smartphones}.
\newblock \emph{National Center for Biotechnology Information (PMC)}, 2023.
\newblock URL \url{https://pmc.ncbi.nlm.nih.gov/articles/PMC10244611/}.

\bibitem[Lewis et~al.(2020)]{lewis2020rag}
P.~Lewis et~al.
\newblock {Retrieval-Augmented Generation for Knowledge-Intensive NLP Tasks}.
\newblock In \emph{Advances in Neural Information Processing Systems (NeurIPS 2020)}, 2020.
\newblock URL \url{https://arxiv.org/abs/2005.11401}.

\bibitem[Hu et~al.(2022)]{hu2022lora}
E.~J. Hu et~al.
\newblock {LoRA: Low-Rank Adaptation of Large Language Models}.
\newblock In \emph{International Conference on Learning Representations (ICLR)}, 2022.
\newblock URL \url{https://arxiv.org/abs/2106.09685}.

\bibitem[Han et~al.(2024)]{han2024peft}
Z.~Han et~al.
\newblock {Parameter-Efficient Fine-Tuning for Large Models: A Comprehensive Survey}.
\newblock \emph{arXiv preprint arXiv:2403.14608}, 2024.
\newblock URL \url{https://arxiv.org/abs/2403.14608}.

\bibitem[{COLING}(2025)]{coling2025chunking}
{COLING}.
\newblock {Mix-of-Granularity: Optimize the Chunking Granularity for Retrieval-Augmented Generation}.
\newblock In \emph{Proceedings of COLING 2025}, 2025.
\newblock URL \url{https://aclanthology.org/2025.coling-main.384/}.

\bibitem[{EMNLP}(2025)]{emnlp2025faithfulness}
{EMNLP}.
\newblock {Benchmarking LLM Faithfulness in RAG with Evolving Leaderboards}.
\newblock In \emph{Proceedings of EMNLP 2025}, 2025.
\newblock URL \url{https://aclanthology.org/2025.emnlp-industry.54/}.

\bibitem[Luo et~al.(2024)]{luo2024causal}
H.~Luo et~al.
\newblock {A Survey on Causal Inference for Recommendation}.
\newblock \emph{arXiv preprint arXiv:2303.11666v2}, 2024.
\newblock URL \url{https://arxiv.org/html/2303.11666v2}.

\bibitem[Jannach et~al.(2023)]{jannach2023multi}
D.~Jannach et~al.
\newblock {A Survey on Multi-Objective Recommender Systems}.
\newblock \emph{National Center for Biotechnology Information (PMC)}, 2023.
\newblock URL \url{https://pmc.ncbi.nlm.nih.gov/articles/PMC10073543/}.

\bibitem[Susser et~al.(2019)]{susser2019manipulation}
D.~Susser et~al.
\newblock {Technology, Autonomy, and Manipulation}.
\newblock \emph{Internet Policy Review}, 2019.
\newblock URL \url{https://policyreview.info/articles/analysis/technology-autonomy-and-manipulation}.

\bibitem[Mathur et~al.(2021)]{mathur2021dark}
A.~Mathur, M.~Kshirsagar, and J.~Mayer.
\newblock {What Makes a Dark Pattern... Dark? Design Attributes, Normative Considerations, and Measurement Methods}.
\newblock In \emph{Proceedings of the 2021 CHI Conference on Human Factors in Computing Systems}, 2021.
\newblock URL \url{https://dl.acm.org/doi/abs/10.1145/3411764.3445610}.

\bibitem[Wang et~al.(2023)]{wang2023fairness}
Y.~Wang et~al.
\newblock {A Survey on the Fairness of Recommender Systems}.
\newblock \emph{ACM Transactions on Information Systems}, 2023.
\newblock URL \url{https://dl.acm.org/doi/10.1145/3547333}.

\bibitem[Fabbri(2023)]{fabbri2023ethical}
M.~Fabbri.
\newblock {An Ethical Perspective on the Implementation of the Transparency Requirements for Recommender Systems Set by the Digital Services Act of the European Union}.
\newblock In \emph{Proceedings of the 2023 AAAI/ACM Conference on AI, Ethics, and Society}, 2023.
\newblock URL \url{https://dl.acm.org/doi/abs/10.1145/3600211.3604717}.

\bibitem[Vaswani et~al.(2017)]{vaswani2017attention}
A.~Vaswani et~al.
\newblock {Attention Is All You Need}.
\newblock In \emph{Advances in Neural Information Processing Systems (NeurIPS 2017)}, 2017.
\newblock URL \url{https://arxiv.org/abs/1706.03762}.

\bibitem[Ho et~al.(2018)]{ho2018notifying}
B.~J. Ho et~al.
\newblock {Notifying Users at the Right Time Using Reinforcement Learning}.
\newblock \emph{Proceedings of the ACM on Interactive, Mobile, Wearable and Ubiquitous Technologies}, 2018.
\newblock URL \url{https://dl.acm.org/doi/pdf/10.1145/3267305.3274107}.

\bibitem[Joachims et~al.(2018)]{joachims2018deep}
T.~Joachims, A.~Swaminathan, and M.~de~Rijke.
\newblock {Deep Learning with Logged Bandit Feedback}.
\newblock In \emph{International Conference on Learning Representations (ICLR)}, 2018.

\end{thebibliography}
\end{document}